\documentstyle[epsfig,longtable]{aipproc}

\font\tenbg=cmmib10 at 10pt

\def \rvecphi{{\hbox{\tenbg\char'036}}}

\begin{document}
\title{Poynting Jets from Accretion Disks}

\author{
R.V.E. Lovelace$^*$,
H. Li$^{\dagger}$,
G.V. Ustyugova$^{+}$,
 M.M. Romanova$^*$,
 and
S.A. Colgate$^{\dagger}$}
    \address{$^*$Department of Astronomy,
Cornell University, Ithaca, NY 14853\\
$^{\dagger}$Theoretical Astrophysics, T-6, MS B288,
Los Alamos National Laboratory, Los Alamos, NM 87545\\
$^+$ Keldysh Institute of
Applied Mathematics, Russian Academy of
Sciences, Miusskaya Square 4, 
Moscow 125047, Russia}

\maketitle

\begin{abstract}

    The powerful narrow
jets observed to emanate from
many compact accreting objects may arise from 
the twisting of a magnetic field
threading a differentially rotating accretion
disk which acts to magnetically 
extract angular momentum and energy from
the  disk.  
   Two main regimes have been discussed,
{\it hydromagnetic outflows}, 
which have a significant mass flux and have
energy and angular momentum 
carried by both the  matter and the
electromagnetic field  and, 
{\it Poynting  outflows}, where the
mass flux is negligible and 
energy and angular momentum are
carried predominantly by the
electromagnetic field. 
    Here we consider a Keplerian
disk initially threaded by
a dipole-like magnetic field and
we present solutions of the force-free
Grad-Shafranov equation for the coronal
plasma .
    We find solutions with Poynting jets 
where there is a continuous outflow
of energy and toroidal magnetic
flux from the disk into the external
space.  
   This behavior contradicts
the commonly accepted ``theorem'' 
of Solar plasma physics
that the motion
of the footpoints of a magnetic
loop structure leads
to a stationary magnetic
field configuration 
with zero power  and flux outflows.

   In addition we discuss recent
magnetohydrodynamic (MHD) simulations 
which establish
that  quasi-stationary collimated
Poynting jets similar to our Grad-Shafranov
solutions arise from the inner
part of a disk threaded by a dipole-like
magnetic field.  
   At the same
time we find that there is  a steady
uncollimated hydromagnetic outflow
from the outer part of the disk.
   The Poynting jets represent a likely model
for the jets from active galactic
nuclei, microquasars, and gamma ray burst sources.

\end{abstract}

\section*{Introduction}

Highly-collimated,
oppositely directed jets are
observed in
active galaxies and 
quasars [1],
and in
old compact stars in binaries 
[2,3].
   Further, well collimated
emission line jets are
seen in young stellar
objects [4,5].
    Different 
ideas and models have been put
forward to explain astrophysical jets
[6] with recent work favoring
 models where twisting of an
ordered magnetic field
threading an accretion
disk acts to magnetically
accelerate the jets [7].
   There are two regimes:
 (1)  the {\it hydromagnetic regime}, 
where energy and angular 
momentum are carried by both
the electromagnetic field and
the kinetic flux of matter, which 
is relevant to the jets from 
young stellar objects;
   and (2) the {\it Poynting flux regime}.
where  energy and angular
from the disk are carried predominantly by the
electromagnetic field, which is relevant
to extra galactic and microquasar jets.

Here,  we discuss recent
theoretical and simulation
work on Poynting jets.   
    Stationary 
Poynting flux dominated
jets have been found in
axisymmetric MHD simulations of
the opening of  magnetic loops 
threading a Keplerian disk [8, 9].
    Theoretical studies 
have developed 
models for Poynting jets 
from accretion disks [10-12].

\section*{Theory of Poynting Outflows}

    Consider
the  coronal magnetic field
of a differentially rotating Keplerian
accretion disk for a given poloidal field
threading the disk.
  The disk is perfectly conducting
with a very small accretion speed.
    Further, consider ``coronal''
or force-free magnetic fields in
the non-relativistic limit.
   Cylindrical $(r,\phi,z)$
coordinates are used and 
axisymmetry is assumed
   Thus the magnetic field has
the form
$ {\bf B}~ = {\bf B}_p +
B_\phi \hat{\rvecphi~}~,
$
with
$
{\bf B}_p = B_r{\hat{\bf r}}+
B_z \hat{\bf z}~.
$
We can write
$
B_r =
-(1 / r)(\partial \Psi
/ \partial z),
B_z =(1 / r)
(\partial \Psi / \partial r),
$
where $\Psi(r,z) \equiv r A_\phi(r,z)$.
   In the 
force-free limit,
the magnetic energy density ${\bf B}^2/(8\pi)$
is much larger than the kinetic or thermal
energy densities; that is, the flow speeds are
sub-Alfv\'enic,
${\bf v}^2 \ll {v}_A^2 = {\bf B}^2/4\pi \rho$,
where $v_A$ is the Alfv\'en velocity.
In this limit,
$
{ 0} \approx {\bf J \times B}$;
   therefore, ${\bf J} = \lambda {\bf B}$
[13] and
   consequently,
\begin{equation}
\Delta^\star \Psi = -
H(\Psi) {d H(\Psi) \over d\Psi}~,
\end{equation}
with $ \Delta^\star \equiv
{\partial^2 / \partial r^2}
-(1/r)(\partial / \partial r)
+{\partial^2 / \partial z^2},
$
which is the Grad-Shafranov equation for
$\Psi$ (see e.g. [10]).

     We consider
an {\it initial value problem} where the
disk at $t=0$ is threaded by a 
dipole-like poloidal magnetic field.
     The form of $H(\Psi)$
is then determined
by the differential rotation of the
disk:
   The azimuthal {\it twist} of a given field
line going from an inner footpoint
at $r_1$ to an outer footpoint at $r_2$
is fixed by the differential rotation
of the disk.
  The field line slippage speed through
the disk due to the disk's finite
magnetic diffusivity is estimated
to be negligible compared with the
Keplerian velocity.
  For a given field line
we have $rd\phi/B_\phi = ds_p/B_p$,
where $ds_p \equiv \sqrt{dr^2+dz^2}$ is the
poloidal arc length along the field
line, and
$B_p \equiv \sqrt{B_r^2+B_z^2}$.
   The total twist of a field line
loop is
\begin{equation}
\label{j31}
\Delta \phi(\Psi) =
\int_1^2 ds_p ~{-B_\phi \over r B_p}
=-H(\Psi) \int_1^2 {ds_p \over r^2 B_p}~,
\end{equation}
with the sign included to give
$\Delta \phi >0$.
  For a Keplerian disk around
an object of mass $M$ the angular
rotation is $\omega_K = \sqrt{G M/r^3}$
so that
the field line twist after a time $t$ is
\begin{equation}
\label{j32}
\Delta \phi(\Psi)
=   \omega_0 ~t
\left[\left({r_0\over r_1}\right)^{3/2} -
\left({r_0\over r_2}\right)^{3/2}\right]
= (\omega_0~t) ~{\cal F}(\Psi/\Psi_0)
\end{equation}
where $r_0$ is the radius of the
$O-$point,
$\omega_0\equiv\sqrt{GM/r_0^3}$,
and ${\cal F}$ is a dimensionless
function (the quantity in the
square brackets).

  The Grad-Shafranov equation (1)
can be readily solved by the method
of successive over-relaxation with
$H(\Psi)$ determined by iteratively
solving equations (1) - (3) (see [12]).
   Figure 1 shows a sample solution
exhibiting a Poynting jet.
  The  solution consists
of a region near the axis which
is {\it magnetically collimated} by
the toroidal $B_\phi$ field and
a region far from the axis, on
the outer radial boundary, which
is {\it anti-collimated} in the sense
that it is pushed against the outer
boundary.
  The field lines returning
to the disk at $r>r_0$ are
anti-collimated by the pressure
of the toroidal magnetic field.

\begin{figure}[b!] 
\centerline{\epsfig{file=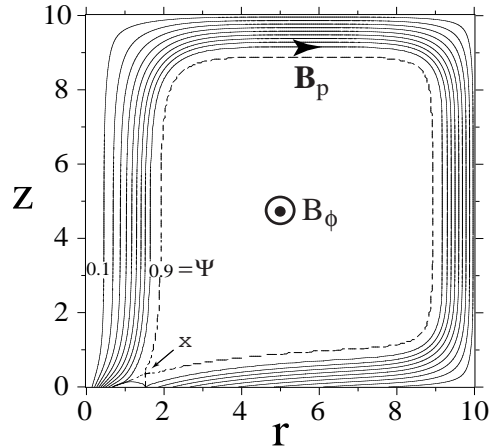,height=2.5in,width=2.5in}}
\vspace{5pt}
\caption{
Poloidal field lines  for Poynting jet
case  for twist
$\omega_0 t =1.84$ rad. and $(-H)_{max} =1.13\Psi_0/r_0$
with $\Psi=$const contours measured in
units of $\Psi_0$ which is the maximum
value of $\Psi$.
  The outer boundaries at $r=R_{m}$
and $z=Z_{m}$ are perfectly conducting
and correspond to an external plasma.
   This external plasma will
expand in response to the magnetic
pressure of the jet so that $R_{m}$
and $Z_{m}$ will increase with time.
  The dashed contour is the separatrix
with the X-point indicated.
  The initial dipole-like poloidal
magnetic field is characterized by
the flux function on the disk surface 
$\Psi(r,z=0)=
(r_0^3/2^{3/2})r^2 K[1-(r/r_0-1)^2/81]/(r_0^2/2+r^2)^{3/2}$,
where $r_0$ is the radius
of the $O-$point of the field in the disk.
    In this and subsequent
plots $\Psi$ is measured in units
of $\Psi_0=r_0^2 K/3^{3/2}$.  
Note that $B_z(0) 
\approx 10.4\Psi_0/r_0^2$.
}
\end{figure}

   Most of the twist 
$\Delta \phi$ of a field line of a Poynting
jet occurs
along the jet from $z=0$ to $Z_{m}$.
Because $-r^2 d\phi/H(\Psi) = dz/B_z$,
we have
\begin{equation}
{\Delta \phi(\Psi) \over -H(\Psi)}=
{(\omega_0 t) {\cal F}(\Psi/\Psi_0)
\over -H(\Psi)} \approx
{Z_{m} \over r^2 B_z}~,
\end{equation}
where $r^2 B_z(r,z)$ is evaluated
on the straight part of the jet
at $r=r(\Psi)$.
   In the core of the jet
$\Psi \ll \Psi_0$, we have
${\cal F} \approx 3^{9/8}
(\Psi_0/\Psi)^{3/4}$, and in
this region we can take
$
\Psi=C \Psi_0 (r / r_0)^q$, and
$ H=-{\cal K}
({\Psi_0 / r_0})
({\Psi/ \Psi_0})^s,$
where $C,~q,~{\cal K},$ and $s$ are
dimensionless constants.
   Equation (1) for the straight
part of the jet implies
$q=1/(1-s)$ and $C^{2(1-s)}=
s(1-s)^2 {\cal K}^2/(1-2s)$.
  Thus we find
$s=1/4$ so that $q=4/3$,
$C =[9/32]^{2/3}{\cal K}^{4/3}$,
and
$
{\cal K}=3^{1/8} 4({r_0 \omega_0  t
/ Z_{m}})$.

   In order to have
a Poynting
jet, we find that ${\cal K}$ must be
larger than
 $\approx 0.5$.
    For the case of uniform expansion
of the top boundary, $Z_{m}=V_z t$,
this condition is the same as
$V_z <9.2 (r_0\omega_0)$.
   For  Figure (1),
${\cal K} \approx 0.844$.
  The field components in the
straight part of the jet
are
\begin{equation}
B_\phi =-\sqrt{2}B_z=-\sqrt{2}
\left({3 \over 16}\right)^{1/3} {\cal K}^{4/3}
\left({\Psi_0 \over r_0^2}\right)
\left({r_0 \over r}\right)^{2/3}~.
\end{equation}
   These dependences agree approximately
with those found in numerical
simulations of Poynting jets [9].
  On the disk,
$\Psi \approx 3^{3/2}\Psi_0(r/r_0)^2$
for $r < r_0/3^{3/4}$.
  Using this and the formula
for $\Psi(r)$
gives the relation between the
radius of a field line in the
disk, denoted $r_d$, and its
radius in the jet,
$
{r / r_0} =6.5
({r_d / r_0})^{3/2} {\cal K}^{-1}$.
Thus the power law for $\Psi$ is applicable for
$r_1 < r <  r_2$, where
$r_1 = 6.5
r_0(r_i/r_0)^{3/2}/{\cal K}$ and
$r_2=1.9r_0/{\cal K}$,
with $r_i$  the inner radius
of the disk.
  The outer edge of the Poynting
jet has a transition layer where
the axial field changes from $B_z(r_2)$
to zero while (minus) the toroidal field
increases from $-B_\phi(r_2)$
to $(-H)_{max}/r_2$.
    Using equations (5), which are
only approximate at $r_2$, gives
$(-H)_{max}\approx 1.2{\cal K} \Psi_0/r_0$, 
which agrees approximately
with our Grad-Shafranov solutions.

\section*{MHD Simulations of Poynting Jets}

Full, axisymmetric MHD
simulations of the evolution
of the coronal plasma of a Keplerian
disk initially threaded by
a dipole-like magnetic field 
are shown in Figure 2.
   For these simulations the outer
boundaries at $r=R_{m}$ and
$z=Z_{m}$ were treated as
free boundaries following the
methods of [14].
   These simulations  established
that a quasi-stationary collimated
Poynting jet arises from the inner
part of the disk while a steady
uncollimated hydromagnetic outflow
arises in the outer part of the disk.

\begin{figure}[b!] 
\centerline{\epsfig{file=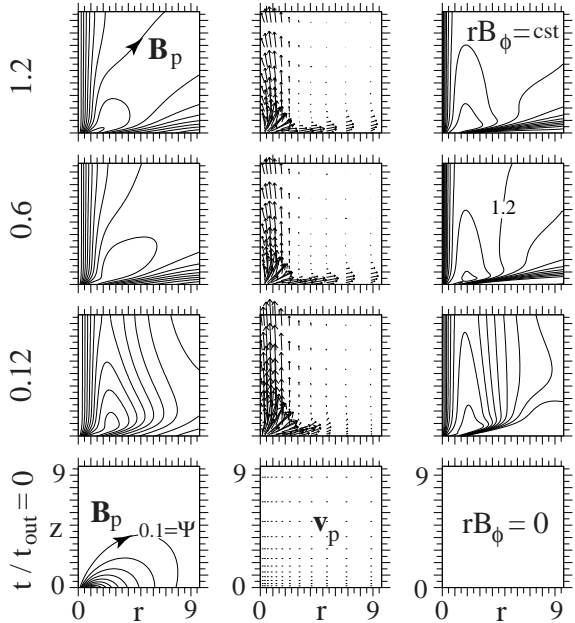,height=3.3in,width=3.in}}
\vspace{5pt}
\caption{
Time evolution of dipole-like
field threading the disk from the
initial configuration $t=0$ (bottom panels)
to the final quasi-stationary 
state $t=1.2 t_{out}$,
where $t_{out}$ is the rotation period of
the disk at the outer radius $R_{m}$ of
the simulation region from [9].  
   The left-hand
panels show the poloidal field lines which
are the same as $\Psi(r,z)=$const lines;
$\Psi$ is
normalized by $\Psi_0$ (the maximum
value of $\Psi(r,z)$) and the
spacing between lines is $0.1$.
  The middle panels show the poloidal
velocity vectors ${\bf v}_p$.
   The right-hand panels show the constant
lines of $-rB_\phi(r,z)>0$ in
units of $\Psi_0/r_0$ and the spacing
between lines is $0.1$.
}
\end{figure}

 Quasi-stationary Poynting jets
 from the two sides of the disk within $r_0$
give an energy outflow
per unit radius of the disk
$d\dot{E}_B/dr=
r v_K(-B_\phi B_z)_h$, where the $h$
subscript indicates evaluation at
the top surface of the disk.
   This outflow
is $\sim r_0d\dot{E}_B/dr \sim
v_K(r_0)(\Psi_0/r_0)^2 \sim  10^{45}({\rm erg/s})
r_{015}^{3/2}\sqrt{M_8}[B_z(0)/6{\rm kG}]^2
$ where  $r_{015}$
is in units of $10^{15}$cm,
and $M_8$ in units of
$10^8{\rm M}_\odot$.
  This formula  agrees
approximately with the values
derived from the simulations.
   The jets also give an
outflow of angular momentum from
the disk which
causes disk accretion
(without viscosity) at the rate
$\dot{M}_B(r) \equiv - 2\pi \Sigma v_r=
-2(r^2/v_K)(B_\phi B_z)_h \sim
2 \Psi_0^2/[r_0^2 v_K(r_0) ]$, where
$\Sigma$ is the disk's surface mass density
[14].
   The Poynting jet has a net axial
momentum flux $\dot{P}_z= (1/4)\int rdr
(B_\phi^2 -B_z^2) \sim
0.5 (\Psi_0/r_0)^2$, which acts to
drive the jet outward through an external
medium.
  Further, the Poynting jet
generates toroidal magnetic flux at the rate
$\dot{\Phi}_t \sim
- 12 [v_K(r_0)/r_0]\Psi_0$.

  For long time-scales,
 the Poynting jet is of course time-dependent
due to the angular momentum it extracts
from the inner  disk ($r<r_0$).
  This loss of angular momentum leads to
a ``global magnetic instability'' and collapse
 of the inner disk [13].
   An approximate model of this collapse
can be made if the
inner disk mass $M_d$ is concentrated
near the $O-$point radius $r_0(t)$,
if the
field line slippage through the
disk is negligible [13],
$\Psi_0=$const, and if
$(-rB_\phi)_{max}\sim \Psi_0/r_0(t)$ (as
found here).  Then,
$
M_d {d r_0 / dt} =  {-2 \Psi_0^2
(GMr_0})^{-1/2}.$
   If  $t_i$ denotes the time at which
$r_0(t_i)=r_i$ (the inner radius of the disk), then
$
r_0(t)=r_i [1 -(t-t_i)/
t_{coll} ]^{2/3}~,
$
for $t \leq t_i$,
where $t_{coll} =\sqrt{GM}~M_d r_i^{3/2}
/(3 \Psi_0^2)$
is the time-scale for
the  collapse of the inner disk.
(Note that the time-scale for $r_0$ to decrease
by a factor of $2$ is $\sim t_i(r_0/r_i)^{3/2}\gg t_i$
for $r_0 \gg r_i$.)
  The power output to the Poynting jets is
$
\dot{E}(t)=(2/ 3)(\Delta E_{tot}/
 t_{coll})[1-(t-t_i)/
 t_{coll}]^{-5/3},$
where $\Delta E_{tot} = G M M_d /2 r_i$ is
the total energy of the outburst.
  Roughly, $t_{coll} \sim 2~{\rm day}
M_8^2(M_d/M_\odot)(6\times 10^{32}{\rm
G cm}^2/\Psi_0)^2$ for
a Schwarzschild black hole, where validity of the
analysis requires $t_{coll} \gg t_i$.
   Such outbursts may explain the flares
of active galactic nuclei blazar sources [16,17]
and the one-time outbursts of gamma ray burst
sources [18].

\section*{Conclusions}

   Recent MHD simulation studies
support the idea that an
ordered magnetic field of
an accretion disk can give
powerful outflows or jets of matter,
energy, and angular momentum.
   Most of the studies  have
been in the hydromagnetic
regime and find
asymptotic flow speeds of the
order of the maximum Keplerian
velocity of the disk, $v_{Ki}$.
   These flows are clearly relevant
to the jets from protostellar systems
which have flow speeds  of the order of $v_{Ki}$.
    In contrast,
observed VLBI jets in quasars and
active galaxies point to bulk
Lorentz factors $\Gamma \sim 10$ - much
larger than the disk Lorentz factor.
   In the jets of gamma ray burst 
sources, $\Gamma \sim 100$.
   The large Lorentz factors
as well as the small Faraday rotation
measures point to the fact that these
jets are in the Poynting flux
regime.
   These jets  may involve
energy extraction
from a rotating black hole 
[19, 20].

    The authors (RVEL, MMR, and SAC) thank
Drs. C. Wheeler and  D.L. Meier
with assistance in attending the
symposium. This research was
supported in part by 
NASA grants NAG5-9047 and NAG5-9735 
and NSF grant AST-9986936.
MMR received partial support
from NSF POWRE grant
AST-9973366.


\begin{references}
\bibitem{be84} Bridle A.H.,
\& Eilek, J.A. (eds) 1984, in
{\it Physics of Energy Transport in
Extragalactic Radio Sources,}
Greenbank: NRAO.

\bibitem{mr94} Mirabel I.F., \&
Rodriguez L.F. 1994
Nature, {\bf 371}, 46.

\bibitem{emm98} Eikenberry S., 
Matthews K., Morgan E.H.,
Remillard R.A., \& Nelson R.W. 
1998, ApJ Lett.,
{\bf 494}, L61.

\bibitem{m85} Mundt R. 1985,
in {\it Protostars and Planets II,}
D.C. Black and M.S. Mathews, eds.
Univ. of Arizona Press, Tucson, 414.

\bibitem{bm88} B\"uhrke T., Mundt R.,
\& Ray T.P. 1988, A\&A, {\bf 200} 99.

\bibitem{} Bisnovatyi-Kogan, G.S. 1993, in 
{\it Stellar Jets and Bipolar Outflows},
eds. L. Errico 
\& A. A. Vittone (Dordrecht: Kluwer), p. 369.

\bibitem{} Meier, D.L., Koide, S., \&
Uchida, Y. 2001, Science, {\bf 291}, 84.


\bibitem{} Romanova M.M.,
Ustyugova G.V., Koldoba A.V.,
Chechetkin V.M., \& Lovelace R.V.E.
1998, ApJ, {\bf 500}, 703.

\bibitem{} Ustyugova G.V.,
Lovelace, R.V.E., Romanova M.M.,
Li, H., \& Cogate, S.A.
2000
ApJ, {\bf 541}, L21.

\bibitem{lws87} Lovelace R.V.E.,
Wang J.C.L.,
\& Sulkanen M.E. 1987,
ApJ, {\bf 315}, 504.

\bibitem{cl98} Colgate S.A. \& Li H. 1998, in
{\it Proc. of VII International Conference and
Lindau Workshop on Plasma Astrophysics
and Space Physics,} Lindau, Germany.

\bibitem{} Li, H., Lovelace, R.V.E.,
Finn, J.M., \& Colgate, S.A. 2001,
in preparation.

\bibitem{gh60} Gold T.,
\& Hoyle F. 1960, Month. Not. R.A.S., {\bf 120}, 7.

\bibitem{ukr99} Ustyugova G.V.,
Koldoba A.V., Romanova M.M.,
Chechetkin V.M., \& Lovelace
R.V.E. 1999
ApJ, {\bf 516}, 221.


\bibitem{lnr97} Lovelace R.V.E.,
Newman W.I., \&
Romanova M.M., 1997, ApJ, {\bf 424}, 628.

\bibitem{rl97} Romanova M.M., \&
Lovelace R.V.E. 1997, ApJ, {\bf 475}, 97.

\bibitem{l98} Levinson A. 1998, ApJ, {\bf 507}, 145.


\bibitem{k97} Katz J.I. 1997, ApJ, {\bf 490}, 633.

\bibitem{bz77} Blandford R.D.,
\& Znajek R.L. 1977,
MNRAS,  {\bf 179}, 433.


\bibitem{lop99} Livio M., Ogilvie G.I., \& Pringle J.E.
1999, ApJ, {\bf 512}, 100.




\end{references}
\end{document}